# A theoretical model for electromagnetic characterization of a spherical dust molecular cloud equilibrium structure

B. Borah and P.K. Karmakar*
*Department of Physics, Tezpur University, Napaam-784028, Tezpur, Assam, India*
**E-mail*: pkk@tezu.ernet.in

**Abstract**
A theoretical model is developed to study the equilibrium electromagnetic properties of a spherically symmetric dust molecular cloud (DMC) structure on the Jeans scale. It applies a technique based on the modified Lane-Emden equation (*m*-LEE). It considers an inhomogeneous distribution of dust grains in field-free hydrodynamic equilibrium configuration within the framework of exact gravito-electrostatic pressure balancing condition. Although weak relative to the massive grains, but finite, the efficacious inertial roles of the thermal species (electrons and ions) are included. A full portrayal of the lowest-order cloud surface boundary (CSB) and associated parameter signatures on the Jeans scale is made numerically for the first time. The multi-order extremization of the *m*-LEE solutions specifies the CSB at a radial point $8.58\times10^{12}$ m relative to the centre. It gets biased negatively due to the interplay of plasma-boundary wall interaction (global) and plasma sheath-sheath coupling (local) processes. The CSB acts as an interfacial transition layer coupling the bounded and unbounded scale-dynamics. The geometrical patterns of the bi-scale plasma coupling are elaborately analyzed. Application of the proposed technique to neutron stars, other observed DMCs and double layers is stressed together with possible future expansion.

## 1 Introduction

The stars and planetary systems are born in turbulent cold self-gravitating interstellar dust molecular clouds (DMCs) (Pandey et al. 1994; Honda 2003; Gao at al. 2010; Klessen et al. 2011; Falco et al. 2013). Star formation in these clouds is governed by complex interplay between the gravitational attraction in the gas and agents such as turbulence, magnetic field, radiation and thermal pressure that resist compression. The supersonic turbulence and thermal instability leads to a transient, clumpy structure. Some of the resulting density fluctuations exceed the critical mass and density of gravitational stability. As these clumps begin to collapse, their central density increases rapidly. Eventually, they give birth to new protostars and other galactic units. Their complex evolution processes result in many electromagnetic phenomena on the new born stars and their atmospheres due to background plasma environment like electromagnetic waves, inductive effects and reconnections (Hale 1913; Rosseland 1924; Gunn 1931; Pandey et al. 1994; Pandey et al. 1996; Verheest 1996; Larson 2003). Their nature is indeed due to collective gravito-electrostatic interplay within the cloud scale (Verheest 1996; Karmakar et al. 2013).

Apart from the DMCs, electromagnetic states, their properties and associated field-induced effects have been discussed by many authors with *electrical stellar models* (ESMs) in past (Hale 1913; Rosseland 1924; Gunn 1931; Ray et al. 2004). The separation of electrical charge inside a star within the ESM framework has been understood by modeling the star as a ball of hot ionized gas (spherical plasma ball) under the light of basic ionization and diffusion processes. Such dynamic processes allow the stellar structure to acquire a net electrical negative charge $\left(Q_s \sim -10^{10}\ C\right)$ on the surface (Rosseland 1924; Gunn 1931). Later, it has been shown that all gravitationally bounded structures possess positive charge, whereas, compensatingly, expanding intergalactic medium acquires negative charge (Bally et al. 1978). But, no work has given any exact solution of their basic structure equations. Some empirical and simple theoretical estimation of the solar and stellar specific surface values of the relevant electromagnetic parameters have merely been provided (Hale 1913; Rosseland 1924; Gunn 1931). The origin mechanism and maintenance of the electric field in such astrophysical situations still remains an open problem to be well studied. Thus, there has been a need of a single self-consistent technique-development for such electromagnetic investigation in equilibrium bounded structures for decades. In addition, the finite inertial character of the plasma thermal species, however small it may be, has never been included in the earlier descriptions.

A polytropic model defined by the Lane-Emden equation (LEE) (Honda et al. 2003; Mirza 2009; Gao et al. 2010) is usually adopted for the description of stellar structure in both force- and mass-balanced conditions under temperature-independent configurations (Milne 1930; Milne 1931; Chandrasekhar 1957; Parand et al. 2008; Bhrawy et al. 2012). There indeed exist various exact solutions for diverse equilibrium configurations describable by the LEE and its various constructs (Khalique et al. 2008). The earlier investigations have ignored the plasma-boundary-wall interaction, gravito-electrostatic coupling processes, and collective scales of the plasma constituents. A full procedural description of the electromagnetic anatomy of the DMCs has been an important challenge for decades from various astrophysical perspectives. What is more especially, it may be noted that there is no model formalism developed so far which gravito-electrostatically couples the self-gravitational contraction (Newtonian dynamics) due to the weight of the massive dust grains, and electrostatic expansion (Coulombic dynamics), resulting from the complicated interaction of the electrically charged grains, to depict electromagnetic behaviour in the inhomogeneous cloud. In addition, the inertial effect of the thermal species on such cloud electrodynamics is still unknown. Therefore, there has been a great need for a long period of time for designing a simple self-consistent technique for investigating the electromagnetic cloud properties of basic interest as a function of collective gravitational weight and electrical charge interaction in presence of active inertial roles of the thermal species. This might systematically be explained on a single potential variable of the cloud, its multi-order derivatives and their extreme behaviour.

In this report, motivated by the importance of basic electromagnetic cloud characterization and its expansion, we propose a simple strategy independent of any polytropic index. The lowest-order inertia-corrected thermal species (Deka et al. 2004; Karmakar et al. 2005; Karmakar et al. 2006; Deka et al. 2010) with all the possible thermal effects, gravito-electrostatic coupling and plasma boundary-wall interaction processes are taken into account in a spherical geometry. We offer a modified LEE ($m$-LEE) scheme (after the self-gravitational Poisson formalism) coupling both the electromagnetic (Hale 1913; Rosseland 1924; Gunn 1931) and hydrostatic (Milne 1931; Chandrasekhar 1957; Avinash et al. 2006; Parand et al. 2008; Bhrawy et al. 2012) behaviors within an integrated gravito-electrostatic framework (Avinash 2006; Avinash et al. 2006; Avinash 2007). The model makes a precise examination whether efficacious inertial contribution of the thermal species affects the existence of the cloud surface boundary (CSB), at least, on the lowest-order, by the balanced gravito-electrostatic interaction, which has earlier been found to be located at a radial point $\xi = 3.50$ on the Jeans scale in like situations (Dwivedi et al. 2007; Karmakar 2010; Karmakar et al. 2011). Efforts are put to see also the detailed electromagnetic aspects on the entire cloud scale, taking care of both force balancing (electromagnetic) and charge balancing (electrostatic) in the fluid form governed by continuity equation (hydrostatic). The different multi-order derivative constructs of the $m$-LEE on the normalized electrostatic pressure considering weak but finite thermal inertia are methodologically obtained. Besides, the derivatives are shown to have important roles in full electromagnetic CSB specification. The electrostatic pressure arises due to the electrostatic repulsion among the shielded dust grains and their inhomogeneous distribution (Avinash 2006; Avinash et al. 2006; Avinash 2007). It is seen that this model is justifiably successful in the cloud characterization of electromagnetic interest with a single dependent variable in the form of the electric pressure only. It offers an extension for detailed characterization of neutron stars, other observed DMCs and double layers in space and astrophysical environments.

Apart from the "Introduction" part already described in section 1 above, this paper is structurally organized in a usual simple format as follows. Section 2 describes physical model setup. Section 3 describes the mathematical formulation. Section 4 presents the results and discussions. Lastly, section 5 depicts the main conclusions along with tentative future applicability through new vistas.

## 2 Physical model

An idealized astrophysical plasma situation of a field-free quasi-neutral self-gravitating cloud consisting of the thermal electrons, ions and inertial dust grains in a spherically symmetric geometry approximation in hydroelectrostatic equilibrium is considered. The assumption of spherical symmetry simplifies the problem (to radial 1-D) mathematically, where, complications due otherwise to multi-order spherical harmonics (in 3-D) and their nonlinear coupling is avoided for now. A bulk equilibrium differential flow is assumed to pre-exist, which is justifiable due to unequal distribution of thermal energies of the heavier and lighter species (i. e., $T_d << T_e \approx T_i = T$ for $m_d >> m_i > m_e$). Global electrical quasi-neutrality is supposed to subsist over the gravito-electrostatically bounded spherical enclosure containing the plasma volume. The solid matter of the assumed identical spherical grains is embedded in the inhomogeneous gaseous phase of the background plasma. We further consider that the heavier grains behave as hydroelectrostatic fluid, whereas, lighter inertia-corrected electrons and ions as the inertia-modified Boltzmannian thermal particles (Deka et al. 2004; Karmakar et al. 2005; Karmakar et al. 2006; Deka et al. 2010) on the Jeans scale. So, the grain self-gravitational interaction would be significant even within the Newtonian point-mass approximation (Hartmann et al. 2001; Grun et al. 2009; Kurgur et al. 2009; Klessen et al. 2011). This means that the grain self-gravity would accelerate the cloud contraction against the Coulombic repulsion. This assumption of thermalization is valid provided the phase velocity of intrinsic background fluctuations, if any, is much smaller than their thermal velocity, i.e., any fluctuation in the electron-ion temperature profile is instantly smoothened out (Pandey et al. 1994; Pandey et al. 1996). In addition, complications like the effects of dispersed grain rotation, kinetic viscosity, non-thermal energy transport (wave dissipation process) and magnetic field due to involved convective circulation dynamics are neglected for simplification. Such simplification would give the simplistic equilibrium picture of the cloud and its average behaviour, particularly, the CSB in absence of any inductive reconnection process (Alfven 1986).

### 3 Mathematical analyses

Our cloud model consists of a charged cloud visualized as a quasi-hydrostatic distribution of the multi-fluid constituent particles. The light neutral gas particles develop a constant background which is weakly coupled to the collapsing charged grains. We describe the model dynamics by the continuity, momentum, and the coupling electro-gravitational Poisson equations with all conventional notations (Pandey et al. 2002; Karmakar et al. 2012) on the astrophysical scales of space and time. The electron and ion dynamics in unnormalized form are described by

$$\frac{\partial n_s}{\partial t} + \nabla.(n_s v_s) = 0\text{, and} \tag{1}$$

$$m_s n_s \left[ \frac{\partial v_s}{\partial t} + (v_s.\nabla) v_s \right] = -q_s n_s \nabla \phi - T_s \nabla n_s . \tag{2}$$

Here, the label $s = (e, i)$ characterizes the electronic and ionic species with charge $q_e = -e$ and $q_i = +e$, respectively. Equations (1)-(2) are the continuity and momentum equations of the flowing electrons and ions with density $n_s$ and velocity $v_s$. The dust dynamics is described by

$$\frac{\partial n_d}{\partial t} + \nabla.(n_d v_d) = 0\text{, and} \tag{3}$$

$$m_d n_d \left[ \frac{\partial v_d}{\partial t} + (v_d.\nabla) v_d \right] = -q_d n_d \nabla \phi - T_d \nabla n_d - m_d n_d \nabla \psi . \tag{4}$$

Again, $n_d$ and $v_d$ represent density and velocity of the dust grains. The spatial distributions of the electrostatic potential $\phi$ and self-gravitational potential $\psi$ in presence of the weak but finite thermal inertia are defined by the closing Poisson equations as follows,

$$\nabla^2 \phi = -4\pi [e(n_i - n_e) - q_d n_d]\text{, and} \tag{5}$$

$$\nabla^2 \psi = 4\pi G (m_d n_d - m_d n_{do} + m_e n_e + m_i n_i), \tag{6}$$

where, $\rho_{d0} = m_d n_{do}$ models the Jeans swindle (Cadez et al. 1990; Vranjes et al. 1994; Verheest et al. 2002; Falco et al. 2013) of the equilibrium unipolar gravitational force field. This swindle provides a formal justification for discarding the unperturbed (zeroth-order) gravitational field and thus, allows us to consider the equilibrium initially as "homogeneous" thereby validating local analysis. Indeed, a spatially homogeneous self-gravitating plasma system cannot be in static equilibrium (for which $\nabla^2\psi \sim 0$), since there is no pressure gradient to balance the gravitational force (originating from the equilibrium cloud-material distribution $\rho_{d0} = m_d n_{do} \sim \left(m_d n_d + m_e n_e + m_i n_i\right)$ as evident from equation (6)). This physically means that self-gravitational potential is sourced only by density fluctuations of the infinite uniform homogeneous background medium. The Jeans assumption (ad hoc) for the self-gravitating uniform homogeneous medium may not be the most suitable one, but it allows us to treat the self-gravitating inhomogeneous plasma dynamics analytically in a simplified way (Vranjes et al. 1994). The results based on this homogenization assumption in most cases have been found to be not far from realistic picture (Cadez et al. 1990; Vranjes et al. 1994; Verheest et al. 2002; Falco et al. 2013). Thus, the full dynamics of the self-gravitating dusty plasma system under consideration is described by the gravito-electrostatically closed set of basic governing equations (1)-(6). In gravito-electrostatic equilibrium of such a cloud, the gravitational pressure ($P_G$) is exactly balanced by the electrostatic pressure ($P_E$) acting radially in opposite directions ($P_G = -P_E$). Therefore, the equation of force balance (Chandrasekhar 1957; Avinash 2006; Avinash et al. 2006; Avinash 2007) of the charged cloud is given by

$$\nabla P_E = -\rho_d \nabla \psi \,. \tag{7}$$

From equation (7) of charged dust state, we obtain a conversion relationship between electric charge density $\left(\rho_E\right)$ and inertial mass density $\left(\rho_d\right)$ in gravito-electrostatic equilibrium as,

$$\rho_d = \Gamma \rho_E \,, \tag{8}$$

where, $\Gamma = \left(4q_d / 3Gm_d\right)$ is a gravito-electrostatic conversion factor. Here, $\Gamma \sim 1$ for $m_d = 3.19 \times 10^{-7}$ kg and $q_d = 100e$ (Avinash 2006; Pandey et al. 1994; Verheest 1996). So, $\rho_d$ is replaced by $\rho_E$ in our calculation scheme for a gravito-electrostatically bounded structure characterization. This type of conversion between charge and mass densities has also been adopted in past to see the effect of charge on charged polytropic compact stars considering their maximum charge accumulation limit (Ray et al. 2004). Applying equations (6)-(8), we obtain the *m*-LEE (Milne 1931; Chandrasekhar 1957; Parand et al. 2008; Avinash et al. 2006; Mirza 2009; Bhrawy et al. 2012) for electrostatic pressure $p_E(r)$ and charge density $\rho_E(r)$ in a spherically symmetric geometry of radius *r* given as follows

$$\frac{1}{r^2}\frac{\partial}{\partial r}\left(\frac{r^2}{\rho_E}\frac{\partial p_E}{\partial r}\right) = -4\pi G \rho_E \,. \tag{9}$$

The modified normalized Boltzmannian population density distributions of the electrons and ions including their weak but finite inertia are obtained from equations (1)-(2) in accordance with the basic rule of inertial drag effects (Deka et al. 2004; Karmakar et al. 2005; Karmakar et al. 2006; Deka et al. 2010), and they are presented as follows,

$$N_e = \exp\left[\left(\frac{1}{2}\right)\frac{m_e}{m_i}M_{eo}^2\{1-\exp(-2\theta)\}+\theta\right], \text{ and} \tag{10}$$

$$N_i = \exp\left[\left(\frac{1}{2}\right)\frac{m_i}{m_d}M_{io}^2\{1-\exp(2\theta)\}-\theta\right]. \tag{11}$$

This is worth mentioning that equation (10) is the lowest-order inertia-corrected population density distribution of the electrons. This means conversely that if we consider $m_e/m_i \to 0$ for the inertialess electrons, equation (10) reduces back to the zero-inertia electrons as given by the normal Boltzmann distribution (Avinash 2006; Karmakar et al. 2012). Similarly, equation (11) is the lowest-order inertia-corrected ion population density.

The electric pressure $P_E$ (Avinash 2006) normalized by the equilibrium plasma thermal pressure $P_{E0} = n_0 T$, ($T_e \cong T_i = T$) with inertia-corrected thermal species (equations (10)-(11)) is derived as follows,

$$P_E = 2\left[-1-\frac{1}{2}\left(\frac{m_i}{m_d}M_{io}^2+\frac{m_e}{m_i}M_{eo}^2\right)\right][\cosh(\theta)-1], \tag{12}$$

where, $\theta = e\phi/T$ is the electrostatic potential developed due to the local charge-imbalance resulting from electrostatically shielded dust-dust repulsive interaction, normalized by the plasma thermal potential $T/e \sim 1\,\text{V}$, for $T` \sim 1\,\text{eV}$ (Avinash 2006; Avinash et al. 2006; Avinash 2007). Also, $N_e = n_e/n_0$, $N_i = n_i/n_0$ and $N_d = n_d/n_0$ are, respectively, the population densities of the electrons, ions and dust grains normalized by the equilibrium plasma population density $n_0$. The electron and ion flow velocities ($M_{eo}$ and $M_{io}$) are normalized by the ion acoustic phase speed ($C_S = (T/m_i)^{1/2}$) and dust sound phase speed ($C_{SS} = (T/m_d)^{1/2}$), respectively. The velocity normalization is such that the weak but finite inertial effect comes into picture on the relevant astrophysical scales of our interest (Deka et al. 2004; Karmakar et al. 2005; Karmakar et al. 2006; Deka et al. 2010).

Now, after weak inertial correction, $\rho_E$ normalized by the equilibrium charge density $\rho_{E0} = n_0 e \sim 1.60\times10^{-12}$ C m$^{-3}$ (for $n_0 = 10^7$ m$^{-3}$ and $e \sim 1.6\times10^{-19}$ C) is given as follows,

$$\rho_E = \left[1+\frac{1}{2}\left(\frac{m_i}{m_d}M_{io}^2+\frac{m_e}{m_i}M_{eo}^2\right)\right]\{2\sinh(\theta)\}+Z_d N_d. \tag{13}$$

Equation (9) in the normalized (with all standard astrophysical parameters) form with all the usual notations can be simplified into the following form,

$$\frac{\alpha}{\xi^2}\frac{\partial}{\partial\xi}\left[\frac{\xi^2}{\rho_E}\left(\frac{\partial P_E}{\partial\xi}\right)\right]=-\rho_E, \tag{14}$$

where, $\alpha = n_o/n_{do} \sim 10^3 - 10^4$ (Avinash 2006). Here, $n_{do}$ is the equilibrium dust population density and $\xi = r/\lambda_J$ is the radial space coordinate normalized by the Jeans scale length $\lambda_J\left(= C_{ss}/\omega_J\right)$. Also, $\omega_J = \left(4\pi G\rho_0\right)^{1/2}$ is the Jeans frequency and $\rho_0$ is the inertia-corrected equilibrium mass density. Equation (14) represents the relationship between the normalized electric pressure $P_E$ and normalized electric charge density $\rho_E$. This may be termed as the electrical analogue of the polytropic *m*-LEE with weak thermal inertia. Now, using equations (12)-(13) in equation (14), we obtain the *m*-LEE on $\theta$-distribution in the reduced form as,

$$\frac{\partial^2\theta}{\partial\xi^2}+A_0\left(\frac{\partial\theta}{\partial\xi}\right)^2+A_1\left(\frac{\partial\theta}{\partial\xi}\right)=-A_2, \tag{15}$$

where, $A_0(\xi)=\dfrac{Z_d N_d \coth(\theta)}{2\,W_1\sinh(\theta)+Z_d N_d}$, $A_1(\xi)=\dfrac{2}{\xi}$, and $A_2(\xi)=\dfrac{1}{\alpha}\dfrac{\left[2\,W_1\sinh(\theta)+Z_d N_d\right]^2}{2W_2\sinh(\theta)}$ with

$$W_1=\left[1+\frac{1}{2}\left(\frac{m_i}{m_d}M_{io}^2+\frac{m_e}{m_i}M_{eo}^2\right)\right] \text{ and } W_2=\left[-1-\frac{1}{2}\left(\frac{m_i}{m_d}M_{io}^2+\frac{m_e}{m_i}M_{eo}^2\right)\right].$$

We are interested in the equilibrium electromagnetic characterization of the spherical cloud. Here, higher-order derivatives of $\theta$ play an important role. Now, equation (15) after spatial differentiation once is transformed to the following simple form,

$$\frac{\partial^3\theta}{\partial\xi^3}+B_0\left(\frac{\partial\theta}{\partial\xi}\right)\left(\frac{\partial^2\theta}{\partial\xi^2}\right)+B_1\frac{\partial^2\theta}{\partial\xi^2}+B_2\left(\frac{\partial\theta}{\partial\xi}\right)^3+B_3\left(\frac{\partial\theta}{\partial\xi}\right)=0, \tag{16}$$

where, $B_0(\xi)=\dfrac{2Z_d N_d\coth(\theta)}{\left[2W_1\sinh(\theta)+Z_d N_d\right]}$, $B_1(\xi)=\dfrac{2}{\xi}$,

$$B_2(\xi)=-\frac{Z_d N_d\left[2W_1\sinh(\theta)\left\{1+\cosh^2(\theta)\right\}+Z_d N_d\right]}{\sinh^2(\theta)\left\{2W_1\sinh(\theta)+Z_d N_d\right\}^2}, \text{ and}$$

$$B_3(\xi)=\frac{\cosh(\theta)\left[4W_1^2-Z_d^2N_d^2\operatorname{cosech}^2(\theta)\right]}{2\alpha W_2}-\frac{2}{\xi^2}.$$

Clearly, equation (16) is the electrostatic version of the third-order *m*-LEE. The various associated coefficients in equation (16) are all functions of the equilibrium plasma parameters.

After spatial differentiation once more, equation (16) gets transformed into the following (4th order) derivative form,

$$\frac{\partial^4\theta}{\partial\xi^4}+C_0\left(\frac{\partial\theta}{\partial\xi}\right)\left(\frac{\partial^3\theta}{\partial\xi^3}\right)+C_1\left(\frac{\partial^3\theta}{\partial\xi^3}\right)+C_2\left(\frac{\partial\theta}{\partial\xi}\right)^2\left(\frac{\partial^2\theta}{\partial\xi^2}\right)+C_3\left(\frac{\partial^2\theta}{\partial\xi^2}\right)^2+C_4\left(\frac{\partial^2\theta}{\partial\xi^2}\right)+C_5\left(\frac{\partial\theta}{\partial\xi}\right)^4$$
$$+C_6\left(\frac{\partial\theta}{\partial\xi}\right)^2+C_7\left(\frac{\partial\theta}{\partial\xi}\right)=0, \tag{17}$$

where, $C_0(\xi)=\frac{2Z_dN_d\coth(\theta)}{[2W_1\sinh(\theta)+Z_dN_d]}$, $C_1(\xi)=\frac{2}{\xi}$,

$$C_2(\xi)=-\frac{5Z_dN_d\left[2W_1\cos ech(\theta)+Z_dN_d\cos ech^2(\theta)+2W_1\coth(\theta)\cosh(\theta)\right]}{[2W_1\sinh(\theta)+Z_dN_d]^2},$$

$$C_3(\xi)=\frac{2Z_dN_d\coth(\theta)}{[2W_1\sinh(\theta)+Z_dN_d]},$$

$$C_4(\xi)=\frac{\left[4W_1^2\cosh(\theta)-Z_d^2N_d^2\cosh(\theta)\cos ech^2(\theta)\right]}{2\alpha W_2}-\frac{4}{\xi^2},$$

$$C_5(\xi)=Z_dN_d[2W_1\sinh(\theta)+Z_dN_d]^{-3}\begin{bmatrix}12W_1^2\coth(\theta)\{1+\cosh^2(\theta)\}-8W_1^2\sinh(\theta)\cosh(\theta)\\-4W_1Z_dN_d\cosh(\theta)+10W_1Z_dN_d\cosh(\theta)\cos ech^2(\theta)\\+2W_1Z_dN_d\sinh(\theta)\coth^3(\theta)+2(Z_dN_d)^2\cosh(\theta)\cos ech^3(\theta)\end{bmatrix},$$

$$C_6(\xi)=\frac{\left[4W_1^2\sinh(\theta)-(Z_dN_d)^2\cos ech(\theta)+2(Z_dN_d)^2\coth^2(\theta)\cos ech(\theta)\right]}{2\alpha W_2},\ \text{and}$$

$$C_7(\xi)=\frac{4}{\xi^3}.$$

If the assumption of spherical symmetry with radial degree of freedom was dropped, then a more realistic three-dimensional (3-D) picture would come into play. Then, we might have to deal with all the components of spherical polar co-ordinates $(r,\theta,\phi)$ in the mathematical formulations. The analytical calculations would become very complex due to the nonlinear coupling of multi-order spherical harmonics. The *m*-LEE might be of different form with different complex coefficients.

However, under the consideration of spherically symmetric simplified configuration, equation (17) represents the electrostatic version of the fourth-order *m*-LEE in differential form. We are interested in the detailed radial profiles of the relevant electromagnetic parameters on the zeroth-order. Being highly complicated, nonlinear and lengthy form, analytical integration for exact solutions is avoided. Applying the fourth-order Runge-Kutta method, it is numerically integrated as an initial value problem to understand the equilibrium structure of the cloud, its lowest-order CSB, associated electromagnetic properties and their transitional behaviors.

## 4 Results and discussions

A theoretical model analysis of the electromagnetic behavior of a simplified self-gravitating spherical charged dust cloud in a field-free inhomogeneous hydroelectrostatic equilibrium configuration is proposed from a new perspective. It involves the application of a new technique derived from the polytropic *m*-LEE on the Jeans scales of space and time. The unique originality is the application of the lowest-order inertial correction (weak, but finite) through the modified Boltzmann distributions of the thermal species in the analytical derivation of the multi-order derivative forms of the *m*-LEE on the normalized electrostatic pressure. A numerical shape-analysis highlights the lowest-order surface characterization even without imposing any conventional polytropic index. Different differentials involved in it typify different electromagnetic significances, thereby characterizing diverse cloud properties. It specifies the lowest-order CSB developed due to plasma sheath-sheath interaction existing around charged dust grains and the plasma boundary-wall interaction effects through gravito-electrostatic coupling. The electromagnetic dynamics evolves accordingly on the bounded interior and unbounded exterior scales. This transitional dynamics is in accord with the earlier predictions by others (Hale 1913; Rosseland 1924; Gunn 1931; Dwivedi et al. 2007, Karmakar 2010; Karmakar et al. 2011).

Before presenting the numerical analyses, different normalization constants are estimated methodologically from the inputs available in the literature (Verheest 1996; Avinash 2006; Avinash et al. 2006; Avinash 2007; Huba 2011). An overview of the normalization constants along with their estimated typical values are tabulated in Table 1 as follows.

**Table 1**. Normalization constants with estimated typical values

| S. No | Physical property | Normalization constant | Typical value |
|---|---|---|---|
| 1 | Distance | Jeans length [ $\lambda_J$ ] | $2.45\times10^{12}$ m |
| 2 | Electrostatic potential | Cloud thermal potential [ $T/e$ ] | $1.00$ V |
| 3 | Electric field | Cloud thermal field [ $T/e\lambda_J$ ] | $4.08\times10^{-13}$ V m$^{-1}$ |
| 4 | Dielectric constant | Permittivity of free space [ $\in_0$ ] | $8.85\times10^{-12}$ F m$^{-1}$ |
| 5 | Electric pressure | Cloud thermal pressure | $1.47\times10^{-36}$ N m$^{-2}$ |
| | | $[\in_0 T^2/e^2\lambda_J^2]$ | |
| 6 | Electric charge density | Equilibrium charge density [ $n_0 e$ ] | $1.60\times10^{-12}$ C m$^{-3}$ |
| 7 | Electric flux | Cloud thermal flux [ $T\lambda_J/e$ ] | $2.45\times10^{12}$ V m |
| 8 | Population densities of electron, ion and grain | Equilibrium plasma population density [ $n_0$ ] | $1.00\times10^{7}$ m$^{-3}$ |
| 9 | Electric energy | Cloud thermal energy [ $\in_0 T^2\lambda_J/e^2$ ] | $2.68\times10^{1}$ J |
| 10 | Inertia-corrected pressure | Thermal pressure [ $n_0 T$ ] | $1.60\times10^{-12}$ N m$^{-2}$ |
| 11 | Ion flow velocity | Dust sound phase speed [ $C_{SS} = (T/m_d)^{1/2}$ ] | $7.08\times10^{-7}$ m s$^{-1}$ |
| 12 | Electron flow velocity | Ion acoustic phase speed [ $C_S = (T/m_i)^{1/2}$ ] | $9.79\times10^{3}$ m s$^{-1}$ |

In order to understand the full electrodynamic cloud features, the multi-order differential form (equation (20)) of the m-LEE is numerically integrated by using the fourth-order Runge-Kutta method with judicious plasma parameter values (Verheest 1996; Avinash 2006; Avinash et al. 2006; Avinash 2007; Huba 2011). The consequent numerical profiles are as shown in Figs. 1-13. Fig. 1 shows the spatial profile of the normalized (by cloud thermal potential) values of (a) Electric potential $(\theta = \theta(\xi))$ (rescaled by dividing with 8 and denoted by blue line), (b) Potential gradient $(\theta_\xi = \partial\theta/\partial\xi)$ (rescaled by multiplying with $10^4$ and denoted by red line), (c) Potential scale length $(L_\theta = [\partial(log\,\theta)/\partial\xi]^{-1})$ (rescaled by multiplying with $3.00\times10^{-3}$ and denoted by green line), and (d) Potential curvature $(\theta_{\xi\xi} = \partial^2\theta/\partial\xi^2)$ (rescaled by multiplying with $3.30\times10^{6}$ and denoted by black line) with normalized (by Jeans scale) position. Various numerical input and initial parameter values adopted are $Z_d = 100$, $N_0 = 1$, $N_d = 2.00\times10^{-3}$, $m_e = 9.11\times10^{-31}$ kg, $m_i = 1.67\times10^{-27}$ kg, $m_d = 1.00\times10^{-13}$ kg, $\alpha = 5.40\times10^{3}$, $(\theta)_i = -3.40\times10^{-3}$, $(\theta_\xi)_i = -1.00\times10^{-10}$, $(\theta_{\xi\xi})_i = -1.00\times10^{-7}$, and $(\theta_{\xi\xi\xi})_i = 1.00\times10^{-11}$. Fig. 2 shows the magnified form of the same potential as in Fig. 1. It is clear that a monotonic potential profile exists in the cloud (Fig. 2) with a value $\theta \sim -3.4005\times10^{-3}$ ($= -3.4005\times10^{-3}$ V) at $\xi = 3.50$. Beyond this point, nonmonotonic features come into picture. The lowest-order CSB by the potential gradient minimization (Fig. 1b) is specified to exist at $r = 3.50\lambda_J = 8.58\times10^{12}$ m for $\lambda_J = 2.45\times10^{12}$ m, calculated with $\rho_0 \sim 10^{-28}$ kg m$^{-3}$ (Avinash 2010). The potential scale-length minimization (Fig. 1c), and zero potential-curature (Fig. 1d) are other characteristic features of the CSB. The normalized potential scale-length at the CSB is $L_{\theta N} \sim -4.67\times10^{-1}$ relative to the centre. This physically is $L_{\theta Phy} = \lambda_J L_{\theta N} \sim -1.15\times10^{12}$ m. By the terminology "lowest-order CSB", we mean the nearest concentric spherical electric potential surface boundary (formed by gravito-electrostatic balancing) relative to the centre of the self-gravitating cloud mass distribution, such that it behaves as an interfacial transition surface exhibiting bounded interior scale (BIS) dynamics on one hand, and unbounded exterior scale (UES) dynamics on the other, as reported earlier in like situations (Dwivedi et al. 2007; Karmakar et al. 2010; Karmakar et al. 2011). Also "curvature" of a parameter here means the "second-order spatial derivative" of the parameter. This is seen that the cloud exhibits perfect quasi-neutrality at the obtained boundary. But, before and after the CSB, an appreciable deviation from quasi-neutrality condition is observed (Fig. 1d). These observations are in good agreement with our earlier results on self-gravitating plasma systems with plasma boundary-wall interaction processes (Dwivedi et al. 2007; Karmakar et al. 2010; Karmakar et al. 2011). As one moves away radially outwards relative to that centre, the cloud may be found to possess next similar higher-order potential (non-rigid) boundaries enclosing the solid grain-matter in the gaseous plasma phase, and so forth, as clearly seen from the curvature profiles. Thus, Figs. 1-2 show that the CSB is not neutral, but it is electrically charged. Applying the normal Coulomb formula (Huba 2011), the total electric charge at the CSB is calculated as $Q\,(=4\pi \in_0 r\theta_{Phy}) \sim -3.24\,\mathrm{C}$ for $\theta_{Phy} \sim -3.4005\times10^{-3}$ V. Fig. 3 displays the profile of the normalized (by cloud thermal field) values of (a) Electric field $(E=-\partial\theta/\partial\xi)$ (blue line), (b) Field divergence $(div\vec{E}=\partial E/\partial\xi)$ (red line), (c) Field scale length $(L_E=[\partial(log\,E)/\partial\xi]^{-1})$ (rescaled by multiplying with $3.00\times10^{-7}$ and denoted by green line), and (d) Field curvature $(E_{\xi\xi}=\partial^2 E/\partial\xi^2)$ (rescaled by multiplying with 5 and denoted by black line) with position. The field monotonically becomes maximum, $E_N=2.19\times10^{-7}$ with the physical value $E_{Phy}(=(T/e\lambda_J)E_N)\sim 8.94\times10^{-20}$ V m$^{-1}$ at $\xi=3.50$. Thereafter, it starts monotonically decreasing to a lower rate ($\sim-1.01\times10^{-7}$, with physical value$\sim-4.12\times10^{-20}$ V m$^{-1}$) at $\xi$~10.00 (Fig. 3a). At $\xi=3.50$, the divergence is exactly zero (Fig. 3b), and the curvature is minimum (Fig. 3d). Thus, the CSB (on the lowest-order) by the electric field maximization, vanishing divergence and curvature minimization simultaneously is re-specified to exist at $\xi=3.50$. Some of us might still question, "What is so specific about $\xi=3.50$?". A simple example for its further clarification is as follows. For a spherical distribution of electric charge (be it conducting, or non-conducting), the electric field gets maximized only at the surface (Sharma et al. 2000). So, conversely, if the electric field gets maximized at some radial point relative to some origin, the point corresponds to the location of the nearest (lowest-order) surface boundary relative to the same origin of the considered charge distribution. We apply the same basic technique of electric field maximization for depicting the lowest-order CSB, found to exist at $\xi=3.50$, which is primarily a field-boundary in our model description. Thus, the cloud has no solid physical (rigid) boundary-wall, but only a diffuse potential boundary (non-rigid) is found to exist. The cloud electric field itself acts as an electrostatic non-rigid wall having variable strengths against self-gravity enclosing the background plasma volume with the maximum strength at the CSB. The field scale length at the CSB is $L_{EN}\sim-2.20\times10^{-1}$ (Fig. 3c), which is physically $L_{EPhys}\sim-5.39\times10^{11}$ m. The corresponding strength of the magnetic field at the CSB is semi-empirically estimated as $B_{Phy}(\sim E_{Phy}/c)\sim 2.98\times10^{-28}$ T, where the light-speed is $c\sim3\times10^{8}$ m s$^{-1}$ (Huba 2011). The magnetic field is too weak to contribute to the cloud dynamics, and hence, neglected at the outset. These findings are found to go in agreement with our earlier results on the self-gravitating plasmas (Dwivedi et al. 2007; Karmakar et al. 2010; Karmakar et al. 2011). After the CSB, field-reversibility occurs on the unbounded scale ($\xi = 7.75$) due possibly to the surface-charge polarization and interstellar radiation-ionization mechanisms. In the field curvature profile (Fig. 3d), nonmonotonic quasi-neutrality deviation is found to exist at near $\xi = 1$, which is due to the thermal pressure driving wave instability followed by compression and rarefaction. The devitation is maximum at the CSB with normalized value $E_{\xi\xi N} \sim -4.00 \times 10^{-8}$, which is physically $E_{\xi\xi Phy} = \left(T/e\lambda_J^3\right)E_{\xi\xi N} \sim -2.72 \times 10^{-45}$ V m$^{-3}$. From the CSB on, monotonic deviation results from the random cloud surface-leakage of the electrons and ions due to their high thermal velocities (Dwivedi et al. 2007; Karmakar et al. 2010; Karmakar et al. 2011). Fig. 4 depicts the profile structure of the normalized (by cloud thermal pressure) values of (a) Electric pressure $\left(P = 1/2\,\chi E^2\right)$ (blue line), (b) Pressure gradient $\left(P_\xi = \partial P/\partial \xi\right)$ (red line), (c) Pressure scale length $\left(L_P = [\partial(log\,P)/\partial\xi]^{-1}\right)$ (rescaled by multiplying with $9.00 \times 10^{-14}$ and denoted by green line), and (d) Pressure curvature $\left(P_{\xi\xi} = \partial^2 P/\partial\xi^2\right)$ (rescaled by multiplying with 2 and denoted by black line) with radial distance. Here, $\chi$ is the dielectric constant of the cloud matter normalized by the permittivity of free space, $\in_0 = 8.85 \times 10^{-12}$ F m$^{-1}$ (Huba 2011). Also $\chi (= \in/\in_0) \sim 1$, since we consider both the media in the same plasma background. The maximum normalized pressure at the CSB comes out to be $P_N \sim 2.38 \times 10^{-14}$ (Fig. 4a), which is physically $P_{Phy} = \left(T^2 \in_0/e^2\lambda_J^2\right)P_N \sim 3.50 \times 10^{-50}$ N m$^{-2}$. The existence of the CSB at $\xi = 3.50$ is further ensured by the joint association of the maximum pressure (Fig. 4a), zero pressure-gradient (Fig. 4b) and minimum pressure-curvature (Fig. 4d). The physical value of the pressure curvature at the CSB is $P_{\xi\xi Phy} \sim -2.15 \times 10^{-63}$ N m$^{-4}$ for $P_{\xi\xi N} \sim -8.75 \times 10^{-15}$. Their small values are due to the application of the small permittivity $\in_0$ and the inertia-modified thermal distributions. The normalized value of the scale length at the CSB is $L_{PN} \sim -1.11 \times 10^{-1}$, which is physically $L_{PPhy} \sim -2.72 \times 10^{11}$ m. The electric pressure is maximum at the CSB due to the electrostatic repulsion between shielded dust grains, repulsion between similar thermal species (like polar) and surface-charge polarization. Moreover, as the strength of the electric field decreases outside the CSB, then, the electrostatic pressure also decreases. Fig. 5 depicts the profile of the normalized (by cloud thermal flux) values of (a) Electric flux $\left(\Phi = 4\pi\xi^2 E\right)$ (rescaled by dividing with 3.50 and denoted by blue line), (b) Flux gradient $\left(\Phi_\xi = \partial\Phi/\partial\xi\right)$ (rescaled by dividing with 3.50 and denoted by red line), (c) Flux scale length $\left(L_\Phi = [\partial(log\,\Phi)/\partial\xi]^{-1}\right)$ (rescaled by multiplying with $3.00 \times 10^{-5}$ and denoted by green line), and (d) Flux curvature $\left(\Phi_{\xi\xi} = \partial^2\Phi/\partial\xi^2\right)$ (black line) with position. The CSB by the zero-flux curvature is further verified to exist at $\xi = 3.50$. The flux at the CSB is $\Phi_N \sim 3.50 \times 10^{-5}$, which is in reality $\Phi_{Phy} = \left(T\lambda_J/e\right)\Phi_N \sim 8.58 \times 10^7$ V m. The flux gradient at the CSB is $\Phi_{\xi Phy} = \left(T/e\right)\Phi_{\xi N} \sim 2.10 \times 10^{-5}$ V for the normalized flux gradient $\Phi_{\xi N} \sim 2.10 \times 10^{-5}$. The electric flux shows variable behaviour with the maximum positive magnitude at near $\xi = 5.5$. The optical thickness (in dimensional form) of the core cloud is given by $\tau = n_d\ \sigma\ z$, where $n_d$ is the population density of dust grain as already mentioned, $\sigma$ is cross-section, and $z$ is the radial thickness of the cloud (Avinash 2007). For low $n_d$ , $\tau$ is small and vice versa. So, beyond the CSB, the equilibrium grain-charge may fluctuate while getting exposed to interstellar UV radiations, which may be responsible for the maximum flux at $\xi = 5.5$. It is interesting to note that the flux shows a bi-polar reversibility at $\xi = 7.5$. According to the integral Gauss law ($\oint \vec{E}.d\vec{S} = \pm Q/\epsilon_0$ ), flux is negative when $E$ and $dS$ are anti-parallel. This happens due to the periodic compression and rarefraction of the plasma species under interstellar radiation-ionization mechanism. The flux scale length at the CSB is $L_{\Phi N} \sim -3.33\times 10^{-1}$, which is physically $L_{\Phi Phy} \sim -8.16\times 10^{11}$ m. Fig. 6 presents the profile of the normalized (by cloud thermal popultion density) values of (a) Electron population density ($N_e$) (rescaled by dividing with 4 and denoted by blue line), (b) Electron population density gradient $\left(N_{e\xi} = \partial N_e/\partial\xi\right)$ (rescaled by multiplying with $1.00\times 10^{7}$ and denoted by red line), (c) Electron population density scale length $\left(L_{N_e} = [\partial(\log N_e)/\partial\xi]^{-1}\right)$ (rescaled by multiplying with $1.00\times 10^{-3}$ and denoted by green line), (d) Ion population density ($N_i$) (rescaled by dividing with 4 and denoted by black line), (e) Ion population density gradient $\left(N_{i\xi} = \partial N_i/\partial\xi\right)$ (multiplying with $1.00\times 10^{7}$ and denoted by cyan line), and (f) Ion population density scale length $\left(L_{N_i} = [\partial(\log N_i)/\partial\xi]^{-1}\right)$ (rescaled by multiplying with $1.00\times 10^{-3}$ and denoted by magenta line) with position. This is seen that the electron and ion densities are slightly different in magnitude (Figs. 6a & 6d), which reveal that the bounded DMC is electrically quasi-neutral on the interior scale. Some extent of non-neutrality may exist on the unbounded scale due to the differential flow motion (thermal) of the plasma constituents. Here, the CSB is re-specified with maximum change of the electron and ion density gradients with opposite polarities. The scale length for the electron and ion population densities are $L_{N_e} = -1.05\times 10^{-7}$ and $L_{N_i} = 9.80\times 10^{2}$, which are physically $L_{N_e Phy} \sim -2.57\times 10^{5}$ m and $L_{N_i Phy} \sim 2.40\times 10^{15}$ m, respectively. Fig. 7 shows the profile of the normalized (by cloud thermal energy) values of (a) Electric energy $\left(U_E = 2/3\pi\chi E^2\xi^3\right)$ (blue line), (b) Energy gradient $\left(U_{E\xi} = \partial U/\partial\xi\right)$ (red line), (c) Energy scale length $\left(L_U = [\partial(\log U_E)/\partial\xi]^{-1}\right)$ (rescaled by multiplying with $4.00\times 10^{-11}$ and denoted by green line), and (d) Energy curvature $\left(U_{E\xi\xi} = \partial^2 U_E/\partial\xi^2\right)$ (black line) with distance. The CSB by the zero-energy curvature (Fig. 7d) is further re-specified to lie at $\xi = 3.50$. The energy at the CSB is $U_{EPhy}\left(\left(=\epsilon_0\, T^2\lambda_J/e^2\right)U_{EN}\right) \sim 9.48\times 10^{-11}$ J for normalized energy $U_{EN} \sim 4.37\times 10^{-12}$. The energy shows variable behaviour with maximum strength on the unbounded scale. As already mentioned, in addition to the CSB, there may be higher-order concentric spherical surface boundaries as well. These boundaries are characterized by the extreme behaviours of the relevant physical parameters beyond the CSB. The basic mechanism for such behaviors may be due to the differential flow of the constituents with differential mass scaling and periodic gravito-acoustic coupling. The scale length at the CSB is $L_{UN} \sim -1.38\times 10^{-1}$, which is physically $L_{UPhy} \sim -3.38\times 10^{11}$ m. It is possible to calculate the magnetic energy $U_{MPhy} = 2/3\pi\mu_0 B^2 r^3$ at the CSB with all usual notations. Taking the plasma permeability as that of vacuum, $\mu_0 = 4\pi\times 10^{-7}$

H m$^{-1}$ (Huba 2011), the CSB magnetic energy comes out to be $U_{MPhy} \sim 1.47\times10^{-22}$ J. Thus, the physical value of the electrostatic-to-magnetic energy ratio is, $U_{EPhy}/U_{MPhy} \sim 10^{11}$. So, various observed phenomena on the Jeans scale are mainly due to the electrical energy transports only, and not due to the magnetic counterpart.

Fig. 8 depicts the profile of the phase portrait between electric potential $(\theta = \theta(\xi))$ and field $(E = -\partial\theta/\partial\xi)$. It shows that the evolution of electric field over the corresponding changes in potential. The field shows a directional reversibity at $\theta = -3.401\times10^{-3}$, a value away from that of the CSB, i.e, in the unbounded scale. Fig. 9, similarly, gives the profile of the phase portrait between electric field $(E = -\partial\theta/\partial\xi)$ and field divergence $(div\vec{E} = \partial E/\partial\xi)$. It likewise shows that the evolution of electric field divergence over the corresponding field value. This is observed that the existence of divergence-free field is one of the stability conditions of the cloud. Again, Fig. 10 exhibits the shape of the phase portrait between electric pressure $(P = 1/2\,\chi E^2)$ and pressure gradient $(P_\xi = \partial P/\partial\xi)$. It displays graphically that the evolution of the electric pressure gradient over the corresponding electric pressure. This geometrical trajectroy is a closed-form curve, and it is within the CSB for $P_N \sim 2.39\times10^{-14}$ at $\xi = 3.50$. The DMC system shows that the centre is the most stable fixed point as the phase trajectroies overlap at that point. Fig. 11 describes the graphical behaviour of the phase portrait between electric energy $(U_E = 2/3\,\pi\chi\,E^2\xi^3)$ and electric energy gradient $(U_{E\xi} = \partial U_E/\partial\xi)$. This is observed that the trajectories of divergence-free field is closely packed and overlap at the centre, which reveals that the centre of the cloud is the most stable fixed point as before. Fig. 12 gives the comparative shape analysis of the normalized (by cloud thermal pressure) values of (a) Electric pressure from the *m*-LEE $(P = 1/2\,\chi\,E^2)$ (rescaled by multiplying with $1.00\times10^5$ and denoted by blue line), (b) Pressure gradient $(P_\xi = \partial P/\partial\xi)$ (rescaled by multiplying with $1.00\times10^5$ and denoted by red line), (c) Electric pressure ($P_E$) with the inertia-correction (rescaled by multiplying with $2.00\times10^{-5}$ and denoted by green line), and (d) Gradient $(P_{E\xi} = \partial P_E/\partial\xi)$ (black line) with position. The CSB location by pressure maximization as well is re-specified at $\xi = 3.50$. Lastly, Fig.13 portrays the same as Fig. 12, but highlighting the microscopic evolution of the inertia-corrected pressure. Thus, the conventional equilibrium pressure (thermal) at the CSB is $P_{CPhy} \sim -1.16\times10^{-5}\,n_0 T = -1.86\times10^{-17}$ N m$^{-2}$. This is estimated for plasma parameter values in interstellar medium with $n_0 = 10^7$ m$^{-3}$ and $T$=1 eV (Avinash 2007; Huba 2011). This is smaller than the corresponding electric pressure obtained from the *m*-LEE. It is clear that both the inertia-corrected pressure and gradients are smaller than those obtained from the *m*-LEE due to the plasma sheath-sheath interaction, and boundary-wall interaction processes.

In addition, one of the most important conjectures derivable from our investigation is that dust acoustic waves and oscillations are prominent within the interior scale of the plasma volume bounded by this CSB (Figs. 1-13). This, however, is not so on the unbounded exterior scale and beyond. Thus, it offers a coarse definition and specification of the lowest-order CSB by the principle of extremization of various relevant electromagnetic parameters, and corresponding transitional dynamics. Weaker electromagnetic parameter values at the CSB, and also beyond, are in qualitative conformity with the existing results previously reported in literature (Hale 1913; Rosseland 1924; Gunn 1931; Dwivedi et al. 2007; Karmakar 2010; Karmakar et al. 2011). This may equally offer an alternate approach to understand the basic physics of the realistic electromagnetic phenomena occurring in self-gravitating objects like stars, clusters and their atmospheres through the proposed *m*-LEE framework. This is because this methodological technique conveniently uses a single self-consistent mathematical construct to depict the entire cloud, its non-rigid boundary, transitional behavior and so forth. Our results are in qualitative agreement as characterized by different spaceprobes, multispace satellite observations and detectors (Hartmann et al. 2001; Grun et al. 2009; Krugur et al. 2009).

We admit that our spherically symmetric model is idealized with full charging of identical spherical grains in the non-relativistic regime. In reality, some deviations may widely exist. For example, it neglects neutral grain dynamics, the dynamical evolution of background gas, spatiotemporal evolution of the inhomogeneous equilibrium, neutral-charged dust interactions, grain-size and grain-mass distributions, etc. Besides, it is developed without the application of any external electromagnetic field. In the derivation of normalized electrical pressure, any contribution due purely to dust-charge fluctuation is absolutely ignored on a time-stationary configuration. All the dissipative agencies are too neglected for simplicity. But recently, many authors have reported like results on the relativistic regime for examining the properties of nonlinear field theories embedded in expanding Euclidean Friedmann-Lemaitre-Robertson-Walker metrics in the context of cosmology and Lagrangian formulation of the stochastic inflationary universe (Escudero 2013; Levasseur 2013). Their path-integral model approach has shared many similarities with the quantum Brownian motion and non-equilibrium statistical quantum formalism under dynamical space-time. The model analysis presented here might be extended to the relativistic limit to study the evolutionary patterns of the DMC electromagnetic properties with the spatial expansion of the universe, especially with a mass-hierarchy, as reported before (Escudero 2013; Levasseur 2013).

It is repeated that the focal aim of the presented analysis is the DMC characterization on the basis of the proposed technique on the astrophysical spatiotemporal scales. Nevertheless, it might be relevant to judge its extensive applicability in diverse other astrophysical situations as well. In case of a neutron star, the surface charge is $Q \sim 10^{20}$ C (Ray et al. 2004). So, its other electromagnetic parameters at the surface boundary are estimated by applying our formalism to test its efficiency. It is seen that different electromagnetic properties of a neutron star can be studied in a simplified way with the input knowledge of its charge only as per the proposed scheme. Likewise, the developed technique can also be extended to other DMCs, such as Barnard 68, 69 and 70; Taurus Molecular Cloud 1 (TMC-1), and Lynds 134N (L134N) and so forth (Alves et al. 2001, Redman et al. 2006). In case of Barnard 68 also with radius $\sim 1.87 \times 10^{15}$ m, the central population density $\sim 2.00 \times 10^{11}$ m$^{-3}$ and dust grain mass $\sim 1.00 \times 10^{-16}$ kg (Alfves et al. 2001, Redman et al. 2006), we obtain the relevant parameter values. In addition to the above, the investigation might have astrophysical applications including the earth's auroral region, extragalactic jets, X-ray and gamma-ray bursts, X-ray pulsars, double radio sources, solar flares, and the source of cosmic ray acceleration like the double layers (Alfven 1986, William 1986). One specific example is the double radio galaxy Cygnus A with electric field strength $\sim 6.20 \times 10^{-2}$ V m$^{-1}$ (Peratt 1996), for which the applicability is examined. The estimated values of diverse electromagnetic parameters for different astrophysical objects by using our strategy are tabulated in Table 2 as follows.

**Table 2.** Estimated values of electromagnetic parameters

| S. No | Parameter | Astrophysical objects | | |
|---|---|---|---|---|
| | | **Neutron star** | **Barnard 68** | **Cygnus A** |
| 1 | Charge ($Q$) | $\sim 10^{20}$ C | $\sim 10^{32}$ C | $\sim 5.07\times10^{14}$ C |
| 2 | Electric potential ($\theta_{phy}$) | $\sim 1.05\times10^{17}$ V | $\sim 1.74\times10^{29}$ V | $\sim 5.32\times10^{11}$ V |
| 3 | Potential scale length ($L_{\theta Phy}$) | $\sim 8.61\times10^{12}$ m | $\sim 8.61\times10^{12}$ m | $\sim 8.61\times10^{12}$ m |
| 4 | Electric field ($E_{Phy}$) | $\sim 1.22\times10^{5}$ V m$^{-1}$ | $\sim 2.03\times10^{16}$ V m$^{-1}$ | $\sim 6.20\times10^{-2}$ V m$^{-1}$ |
| 5 | Field curvature ($E_{\xi\xi Phy}$) | $\sim 1.66\times10^{-21}$ V m$^{-3}$ | $\sim 2.76\times10^{-10}$ V m$^{-3}$ | $\sim 8.42\times10^{-16}$ V m$^{-3}$ |
| 6 | Field scale length ($L_{EPhy}$) | $\sim 8.58\times10^{12}$ m | $\sim 8.58\times10^{12}$ m | $\sim 8.58\times10^{12}$ m |
| 7 | Magnetic field ($B_{Phy}$) | $\sim 4.07\times10^{-5}$ T | $\sim 6.77\times10^{7}$ T | $\sim 2.06\times10^{-10}$ T |
| 8 | Electric pressure ($P_{Phy}$) | $\sim 6.58\times10^{-2}$ N m$^{-2}$ | $\sim 1.82\times10^{21}$ N m$^{-2}$ | $\sim 1.70\times10^{-14}$ N m$^{-2}$ |
| 9 | Pressure curvature ($P_{\xi\xi Phy}$) | $\sim 8.94\times10^{-28}$ N m$^{-4}$ | $\sim 2.47\times10^{-5}$ N m$^{-4}$ | $\sim 2.30\times10^{-40}$ N m$^{-4}$ |
| 10 | Pressure scale length ($L_{PPhy}$) | $\sim 8.58\times10^{12}$ m | $\sim 8.58\times10^{12}$ m | $\sim 8.58\times10^{12}$ m |
| 11 | Electric flux ($\Phi_{Phy}$) | $\sim 1.13\times10^{32}$ V m | $\sim 1.87\times10^{43}$ V m | $\sim 5.73\times10^{13}$ V m |
| 12 | Flux gradient ($\Phi_{\xi Phy}$) | $\sim 1.31\times10^{19}$ V | $\sim 2.17\times10^{30}$ V | $\sim 6.67$ V |
| 13 | Flux curvature ($\Phi_{\xi\xi Phy}$) | $\sim 1.52\times10^{6}$ V m$^{-1}$ | $\sim 2.52\times10^{17}$ V m$^{-1}$ | $\sim 7.77\times10^{-13}$ V m$^{-1}$ |
| 14 | Flux scale length ($L_{\Phi Phy}$) | $\sim 8.58\times10^{12}$ m | $\sim 8.58\times10^{12}$ m | $\sim 8.58\times10^{12}$ m |
| 15 | Electric energy ($U_{EPhy}$) | $\sim 2.03\times10^{37}$ J | $\sim 4.82\times10^{60}$ J | $\sim 5.08\times10^{24}$ J |
| 16 | Energy gradient ($U_{E\xi Phy}$) | $\sim 2.46\times10^{24}$ J m$^{-1}$ | $\sim 5.61\times10^{47}$ J m$^{-1}$ | $\sim 5.92\times10^{11}$ J m$^{-1}$ |
| 17 | Energy curvature ($U_{E\xi\xi Phy}$) | $\sim 2.86\times10^{11}$ J m$^{-2}$ | $\sim 6.53\times10^{24}$ J m$^{-2}$ | $\sim 6.89\times10^{-2}$ J m$^{-2}$ |
| 18 | Energy scale length ($L_{UPhy}$) | $\sim 8.58\times10^{12}$ m | $\sim 8.58\times10^{12}$ m | $\sim 8.58\times10^{12}$ m |
| 19 | Magnetic energy ($U_{MPhy}$) | $\sim 3.10\times10^{24}$ J | $\sim 7.38\times10^{48}$ J | $\sim 7.07\times10^{13}$ J |
| 20 | Electric-to-magnetic energy ratio ($U_{EPhy}/U_{MPhy}$) | $\sim 10^{11}$ | $\sim 10^{11}$ | $\sim 10^{11}$ |

In case of plane monochromatic electromagnetic wave propagation through vacuum, both electric and magnetic energy densities are equal. So, for a particular volume in vacuum, the corresponding energies are also equal (Griffiths 2008). In contrast, our model deals neither with purely plane monochromatic wave propagation, nor with vacuum as the propagation medium. An astrophysical plasma medium is considered whose absolute permittivity is still unknown (Huba 2011). Subsequently, we analyze the medium with the help of absolute permittivity of free space, the best possible choice (Huba 2011). So, some deviations are found to occur on the parametric quantification, thereby giving $U_{EPhy}/U_{MPhy} \sim 10^{11}$, which is a constant throughout the astrophysical situations (Table 2). Thus, from all the above examples, it is found that the ratio of electric-to-magnetic energy is a constant. It hereby shows that the electric force is more dominating than the magnetic counterpart in the context of astrophysical ionized matter-matter interactions on the Jeans scales of space and time. This extrapolation goes in good correspondence with the earlier predictions on astrophysical electromagnetism (Bally et al. 1978) hand in hand.

## 5 Conclusions

In this paper a methodological study of the relevant electromagnetic properties of a self-gravitating spherical DMC is carried out on the astrophysical scale. The gravito-electrostatic equilibrium structure of the cloud is modeled using analytical, graphical and numerical techniques. It applies the *m*-LEE formalism based on hydroelectrostatic polytrope. The effects of the lowest-order inertial correction of the thermal electrons and ions are taken into account amid diverse spatial inhomogeneities. The basic framework of the *m*-LEE calculation scheme is based on a coupling of the Newtonian and Coulombic dynamics of the constituents. The genesis of the proposed technique of the cloud characterization lies in the diverse relevant electromagnetic properties, their multi-order gradients, scale lengths and extreme behaviors on both the bounded and unbounded scales using a single potential variable of electrodynamical significance.

An interesting property on the existence of the lowest-order CSB (at $\xi$ ~3.50 on the Jeans scale) in perfect agreement with the gravito-electrostatic field maximization principle (Dwivedi et al. 2007; Karmakar 2010; Karmakar et al. 2011) is also reported. One of the most important conclusions drawn from this study is that dust acoustic waves and oscillations are more prominent within the plasma volume bounded self-gravitationally by this boundary, but not so beyond that due to fully charged grains. This analysis may form an elementary input to further study of self-gravitating dusty plasma, astrophysical bounded equilibrium structures, and associated various characteristics of electromagnetic origin in more realistic situations. In such situations, in which both self-gravity and dust are important, it may be noted that the use of an average grain-charge (non-fluctuating) approximation may not be so appropriate.

Deviating slightly from the principal aim of the study, we examine the applicability of our model for realistic characterization of neutron stars, other observed DMCs and double layers also together with future expansion possibilities in space and astrophysical environments. It is pertinent to add further that neutral gas, neutral grains, ions, electrons, and charged grains all need to be considered simultaneously along with suitable equations of state for future refinements in our model. To summarize, we repeat the major conclusive remarks based on our simplified model study briefly as follows.

(1) The lowest-order CSB of the spherically symmetric DMC, delineating the bounded interior and unbounded exterior scales, is precisely determined. It adopts a new theoretical technique based on the *m*-LEE calculation scheme in the non-relativistic regime with the inertial correction.

(2) The CSB is found to exist at $\xi = 3.50 = 8.58 \times 10^{12}$ m by the maximization principle of electric field, and also by the extremization of the associated relevant parameters. The radial scale-size of the CSB ($r = 3.50\lambda_J = 8.58 \times 10^{12} \sim 10^{13}$ m) is in exact correspondence with that ($L \sim 10^{13}$ m) of the Avinash-Shukla mass limit ($M_{AS} \sim 10^{25}$ kg) for the stable cloud mass with grain-size effect (Avinash 2010).
(3) The DMC is negatively biased with negligible variation of electrostatic potential from the center to the CSB, and electric field maximizes at the CBS due to the space-charge polarization and sheath-sheath interaction processes resulting from charged grain-grain coupling.
(4) The phase-space portraits for understanding the cloud-stability behaviours in a geometrical pattern from a new outlook of gravito-electrostatic interplay are presented.
(5) The magnitude of electric potential with the inertia-corrected thermal species is found to be very small. The basic physics behind is that due to the inclusion of weak but finite inertia, the dust grains become less polarized, and hence, it lowers the observed value of potential.
(6) The basic physical mechanism responsible for the CSB is the joint action of plasma-wall interaction and plasma sheath-sheath coupling processes amid repulsive charged grains.
(7) The basic properties of the DMC are dominated by the electrical parameters as the strength of magnetic field is negligible ($U_{EPhy} / U_{MPhy} \sim 10^{11}$), which is in agreement with others.
(8) Dust acoustic waves, oscillations and fluctuations are more prominent within the plasma volume bounded by this boundary, thereby supporting admixture of both the Jeans (bounded) and acoustic (unbounded) modes.
(9) The obtained boundary shows a two-scale transitional behavior from an unstable to almost stable one in terms of its normal electromagnetics and their associated scale lengths.
(10) Irrespective of others, this model is successful in the astrophysical DMC characterization with a single dependent variable in the form of the electric pressure. The pressure is due to charged dust-dust interaction (developing polarization field) and density inhomogeneity without using any conventional polytropic index.
(11) Most of the observed phenomena on field-matter interactions in space and astrophysical environments are mainly due to the electrical energy transport only, and not due to the magnetic counterpart idealistically.
(12) Smaller values of various electromagnetic parameters are still debatable for vacuum electric permittivity and magnetic permeability adopted in place of those really unknown for the plasma.
(13) Electric field reversibility (due to the thermal pressure driving dust acoustic wave instability through some compression and rarefaction) is observed on the bounded interior scale.
(14) Hydroelectrostatic equilibrium of the DMC is indeed inhomogeneous in nature.
(15) The net charge at the CSB is estimated as $Q \sim -3.24$ C. But, in contrast, in case of the dust-free gravito-electrostic sheath model analysis (Dwivedi et al. 2007; Karmakar et al. 2011), it comes out to be $Q \sim -1.20 \times 10^{2}$ C. This lowering deviation is due to the loss of the thermal electrons in the charging process of the grains, and the subsequent sheath-sheath interaction developed around each of the shielded grain.
(16) The surface pressure in our model exceeds the inertia-corrected pressure. The main reason is attributable to the dynamical loss of the thermal species in recombination and charging of the grains. Our model pressure is greater due to the plasma boundary-wall interaction processes, sheath-sheath interactions, and charging-sheilding mechanisms of the dust-grains in presence of the weak but finite inertia of the thermal species on the Jeans scales of space and time.
(17) Lastly, although our simplified model is originally developed for the DMC characterization with a single potential parameter, it parallelly provides extensive applications for further study of the electromagnetic state of diverse realistic astrophysical objects, their constituent dust grains of various characteristics, and ambient dusty atmospheres by an exact polytropic sphere even without the help of any conventional kind of typifying polytropic indices.

## Acknowledgements

The valuable comments, specific remarks and precise suggestions by the anonymous referees to refine the scientific contents of the manuscript are gratefully acknowledged. The financial support from the Department of Science and Technology (DST) of New Delhi, Government of India, extended through the SERB Fast Track Project (Grant No. SR/FTP/PS-021/2011) is also thankfully recognized.

**Figures:**

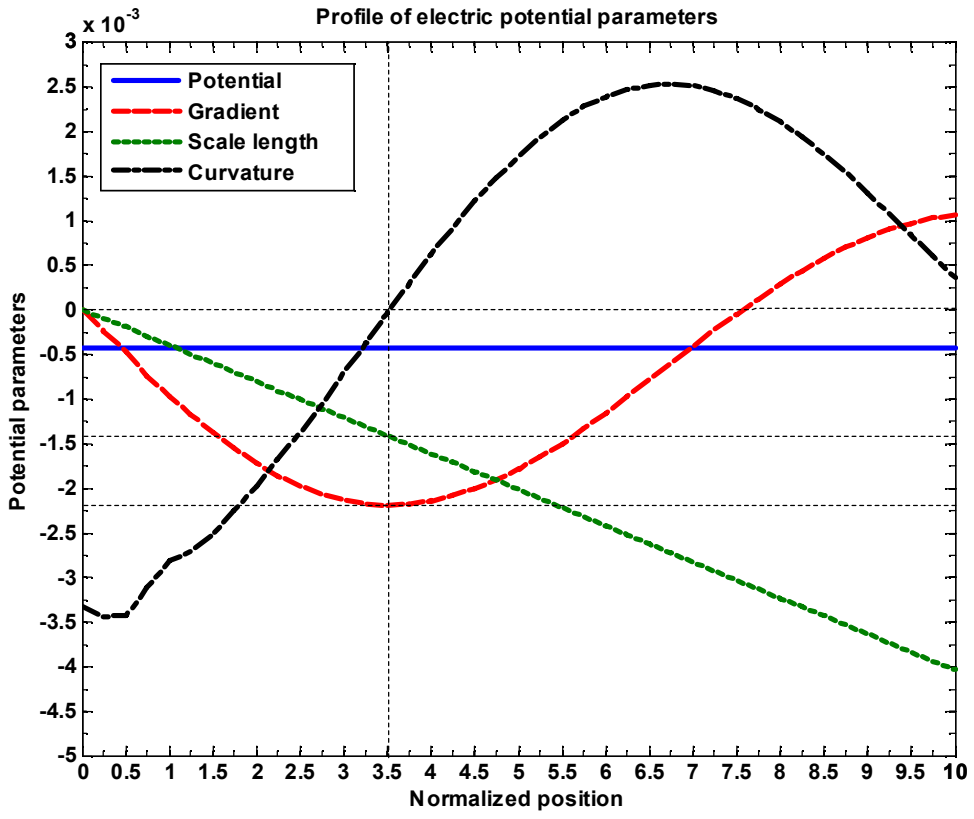

**(1)**

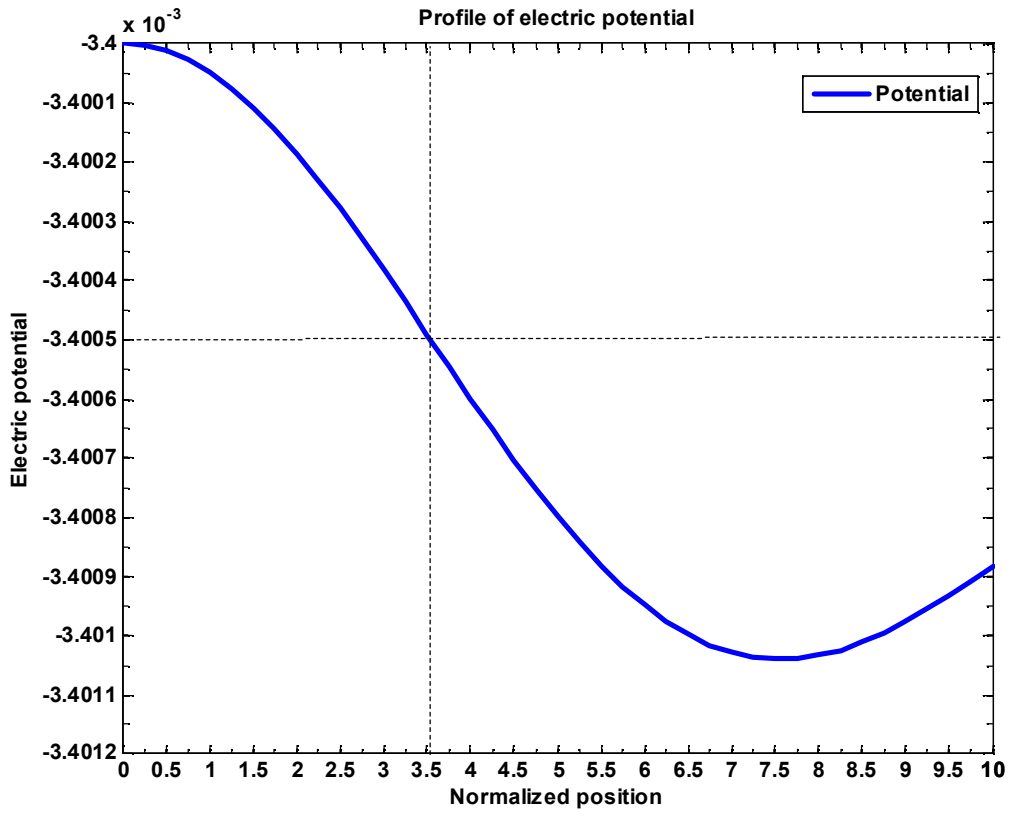

**(2)**

**Figure 1.** Profile of the normalized values of (a) Electric potential (blue line), (b) Potential gradient (red line), (c) Potential scale length (green line), and (d) Potential curvature (black line) with normalized position in rescaled form. Various numerical input and initial parameter values adopted in the simulation are $Z_d = 100$, $N_0 = 1$, $N_d = 2.00 \times 10^{-3}$, $m_e = 9.11 \times 10^{-31}$ kg, $m_i = 1.67 \times 10^{-27}$ kg, $m_d = 1.00 \times 10^{-13}$ kg, $\alpha = 5.40 \times 10^3$, $(\theta)_i = -3.40 \times 10^{-3}$, $(\theta_\xi)_i = -1.00 \times 10^{-10}$, $(\theta_{\xi\xi})_i = -1.00 \times 10^{-7}$ and $(\theta_{\xi\xi\xi})_i = 1.00 \times 10^{-11}$. **Figure 2**. Profile of the normalized values of electric potential with normalized position under the same condition as Fig. 1.

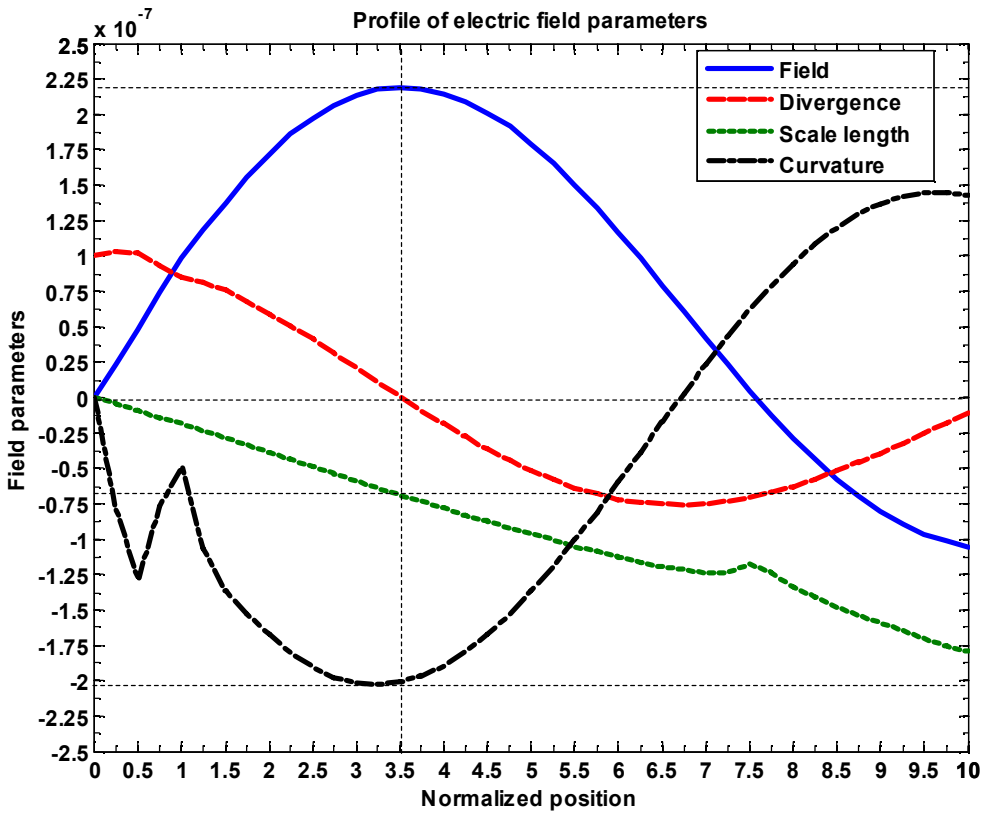

**(3)**

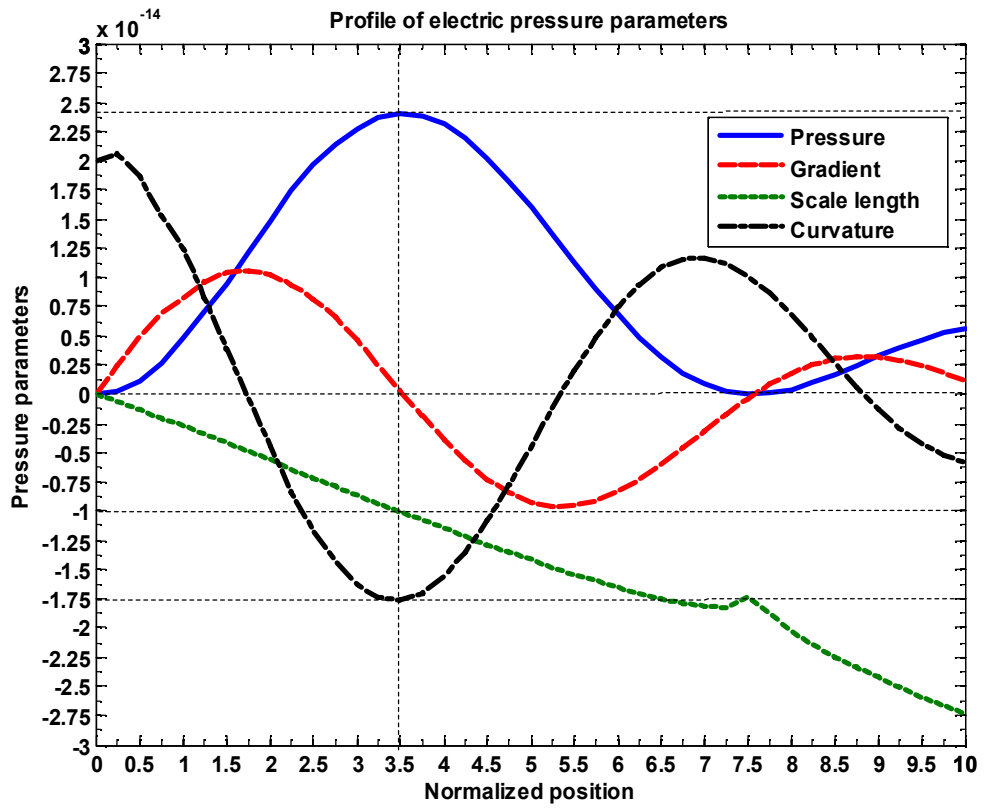

**(4)**

**Figure 3**. Profile of the normalized values of (a) Electric field (blue line), (b) Field divergence (red line), (c) Field scale length (green line), and (d) Field curvature (black line) with normalized position in rescaled form. **Figure 4**. Profile of the normalized values of (a) Electric pressure (blue line), (b) Pressure gradient (red line), (c) Pressure scale length (green line), and (d) Pressure curvature (black line) with normalized position in rescaled form. Various numerical input and initial parameter values are the same as before.

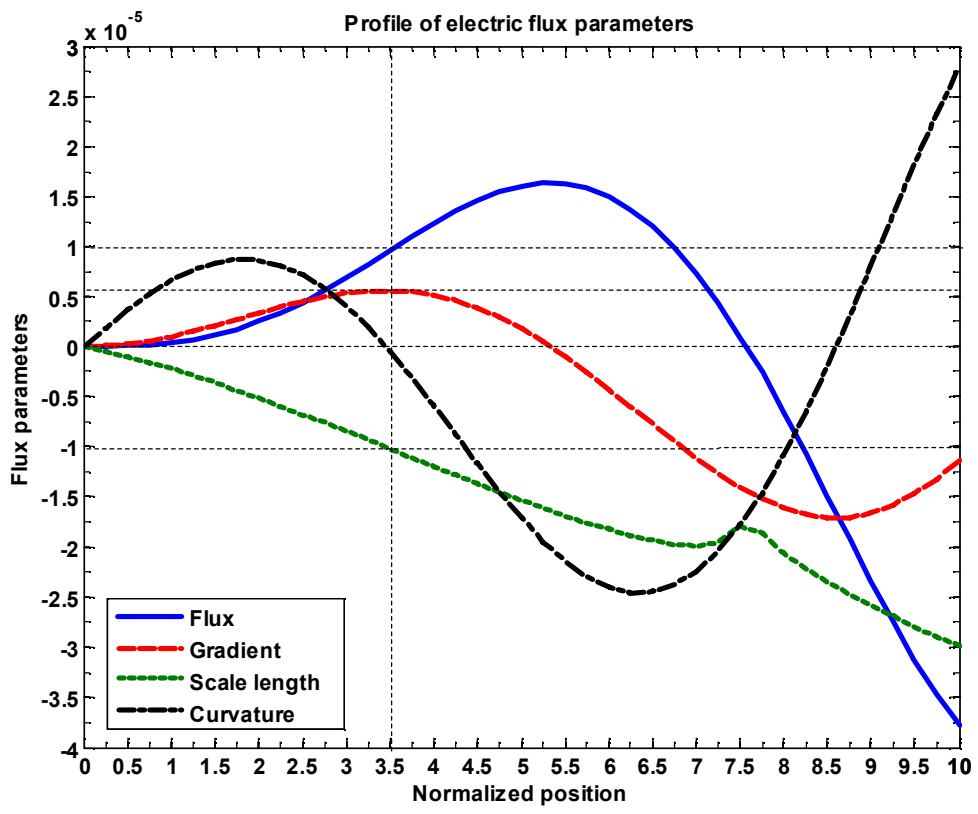

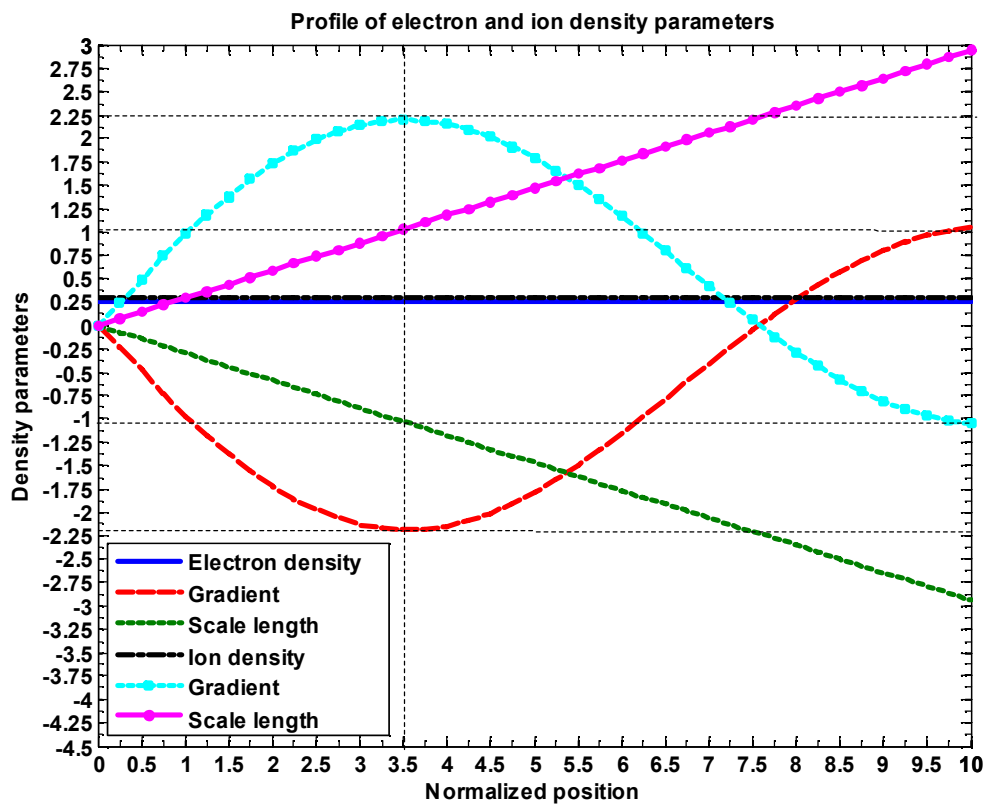

**(5)** **(6)**

**Figure 5**. Profile of the normalized values of (a) Electric flux (blue line), (b) Flux gradient (red line), (c) Flux scale length (green line), and (d) Flux curvature (black line) with normalized position in re-scaled form. **Figure 6**. Profile of the normalized values of (a) Electron population density (blue line), (b) Electron population density gradient (red line), (c) Electron population density scale length (green line), (d) Ion population density (black line), (e) Ion population density gradient (cyan line), and (f) Ion population density scale length (magenda line) with normalized position in rescaled form. Various numerical input and initial parameter values are the same as before.

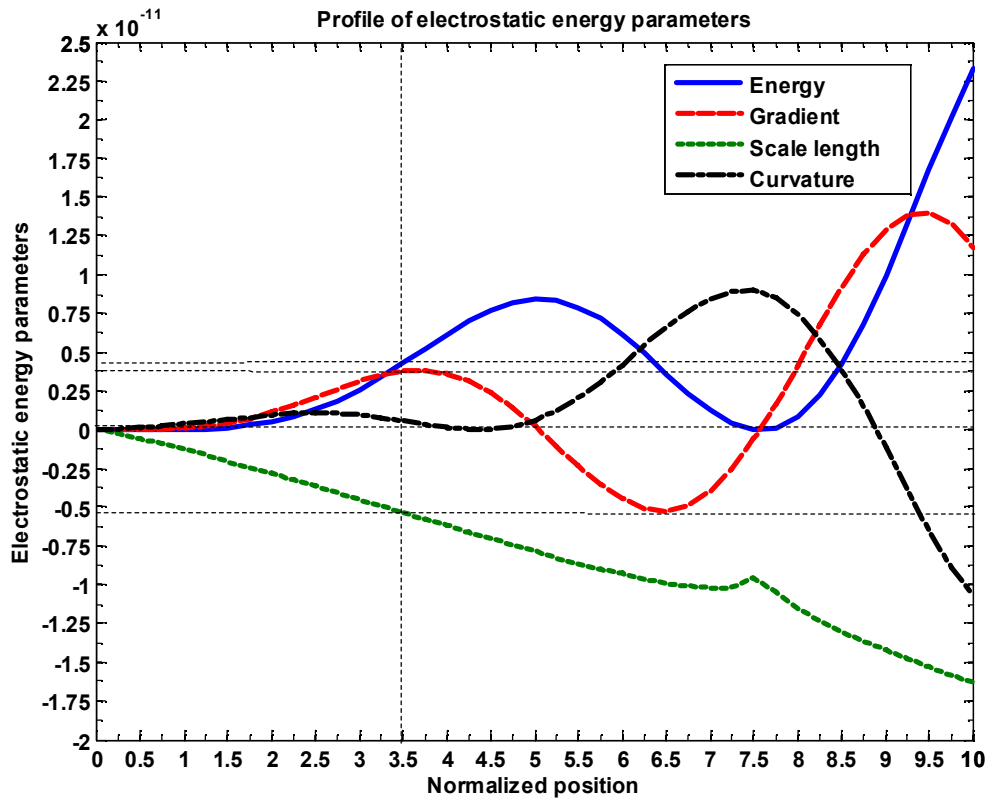

**(7)**

**Figure 7**. Profile of the normalized values of (a) Electric energy (blue line), (b) Energy gradient (red line), (c) Energy scale length (green line), and (d) Energy curvature (black line) with normalized position in rescaled form. Various numerical input and initial parameter values are the same as before.

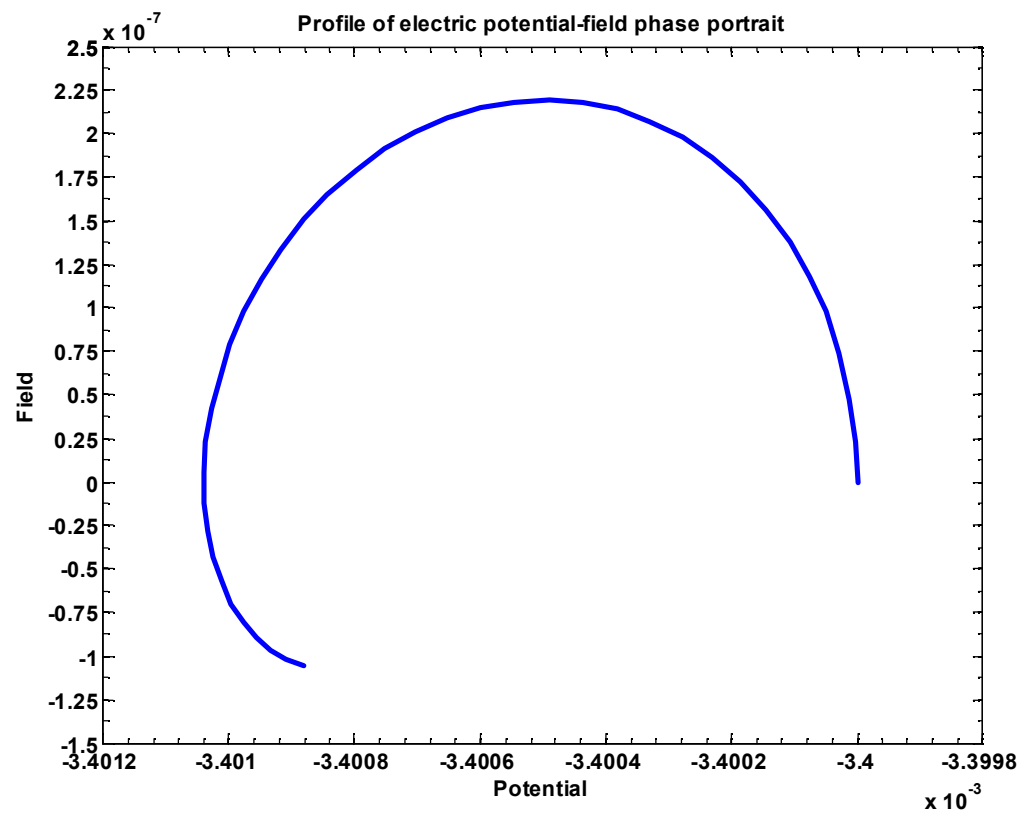

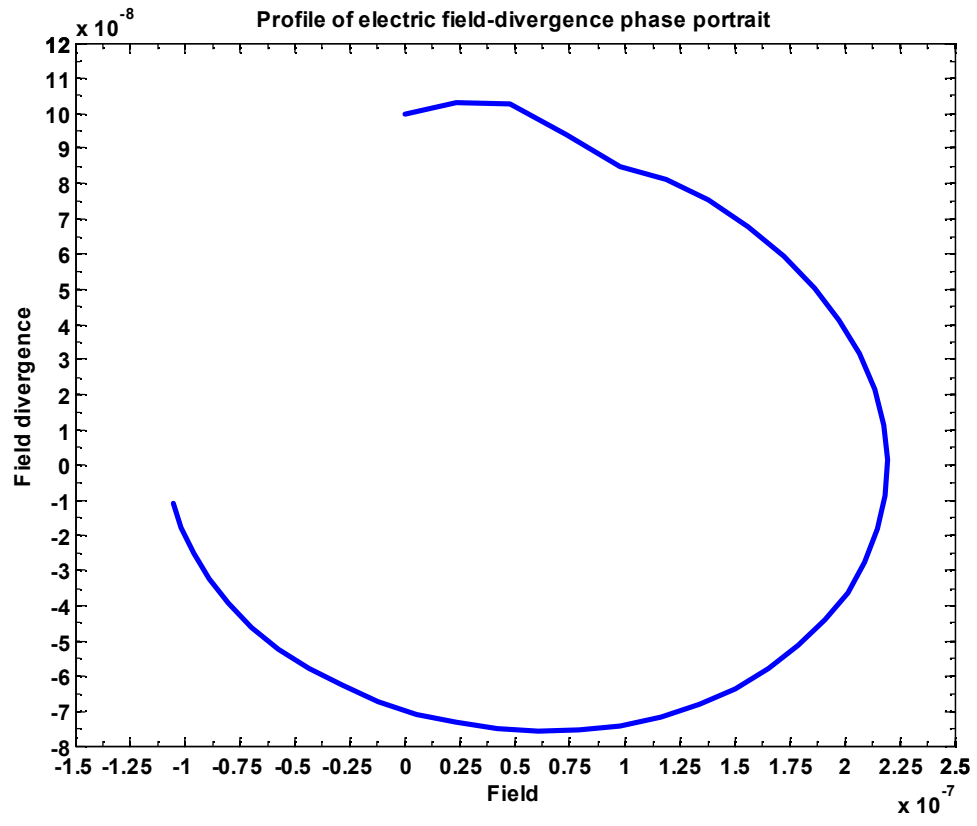

**(8)** **(9)**

**Figure 8**. Profile of phase portrait between electric potential and field. **Figure 9**. Profile of phase portrait between electric field and field divergence. Various numerical input and initial parameter values the same as before.

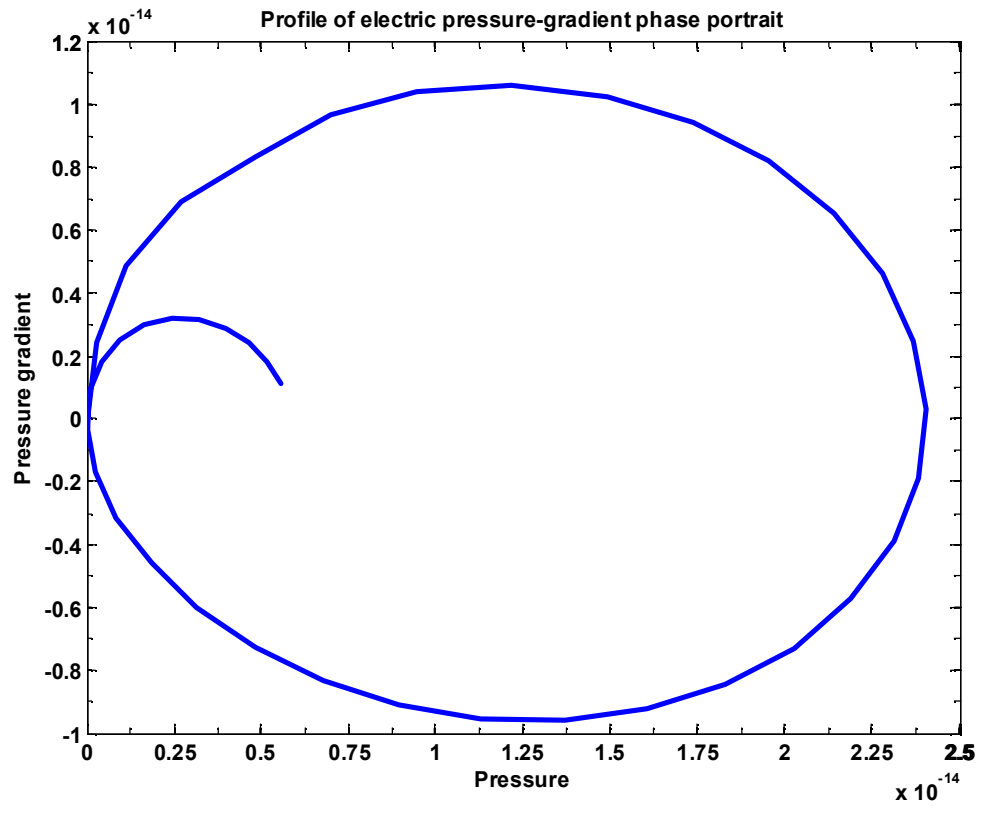

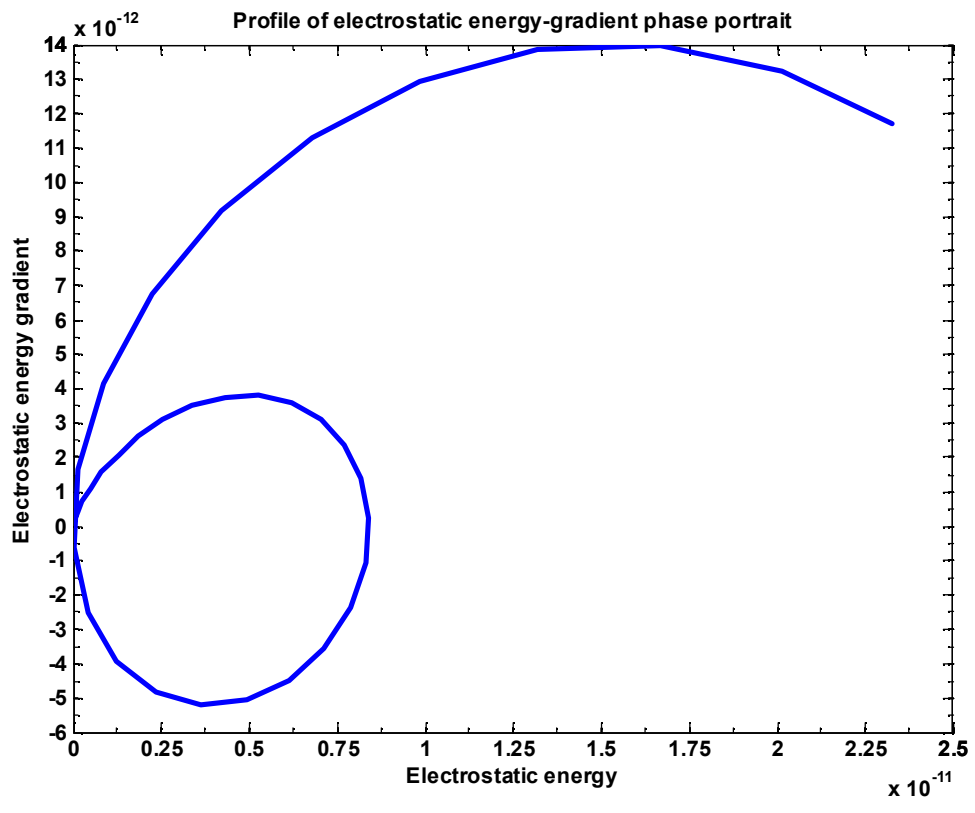

**(10)** **(11)**

**Figure 10**. Profile of phase portrait between electric pressure and pressure gradient. **Figure 11**. Profile of phase portrait between electric energy and electric energy gradient. Various numerical input and initial parameter values are the same as before.

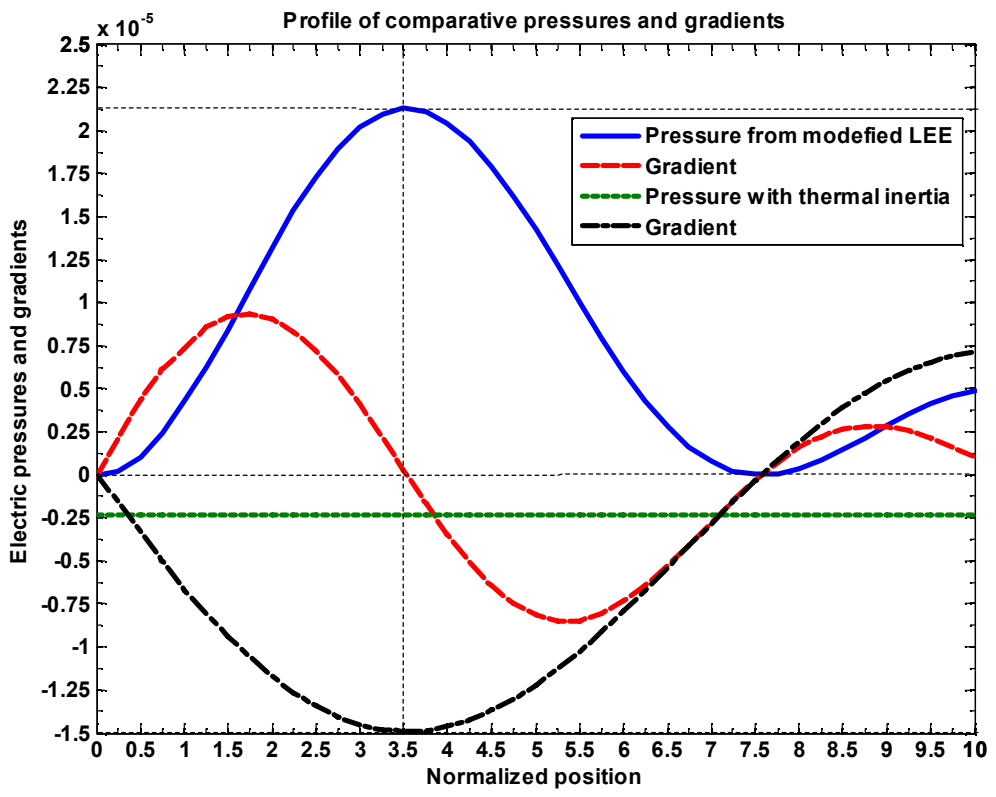

**(12)**

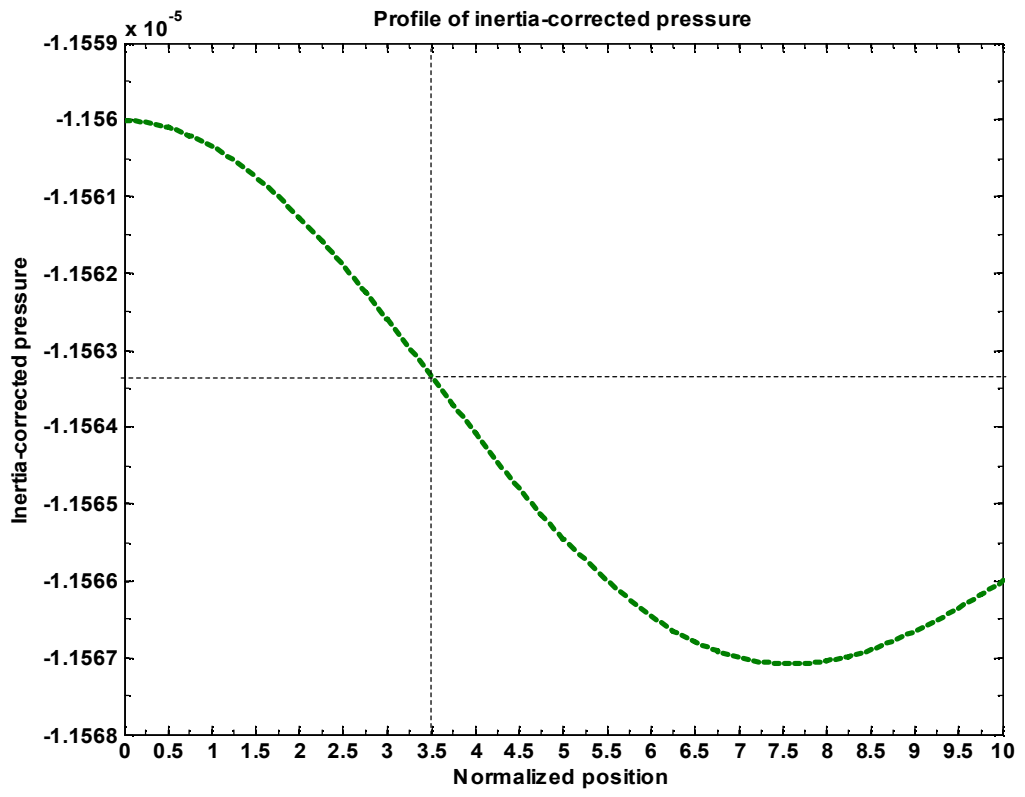

**(13)**

**Figure 12.** Profile of the normalized values of (a) Electric pressure from modefied LEE (blue line), (b) Pressure gradient (red line), (c) Electric pressure with thermal inertia correction (green line), and (d) Gradient (black line) with normalized position in rescaled form. **Figure 13**. Profile of the normalized electric pressure with thermal inertia correction versus normalized position. Various numerical input and initial parameter values are the same as adopted before.